\documentclass[12pt]{article}
\pdfoutput=1
\usepackage[utf8]{inputenc}
\usepackage[T1]{fontenc}
\usepackage{float}
\usepackage{authblk}
\usepackage[normalem]{ulem}
\usepackage{xcolor}
\usepackage{amsmath}
\usepackage{amsthm}
\usepackage{amssymb}
\usepackage{adjustbox}
\usepackage{graphicx}
\usepackage{booktabs}
\usepackage{mathtools}
\usepackage{fullpage}
\usepackage[mathlines]{lineno}
\usepackage{multirow}
\usepackage{subcaption}
\usepackage[style=numeric-comp,sorting=none,maxnames=99]{biblatex}

\usepackage[colorlinks = true,
            linkcolor = teal,
            urlcolor  = teal,
            citecolor = teal]{hyperref}

\title{Spin glass systems as collective active inference}
\author[1-4]{Conor Heins\thanks{\href{mailto:conor.heins@gmail.com}{conor.heins@gmail.com}}}
\author[1,4,5]{Brennan Klein}
\author[4,6]{Daphne Demekas}
\author[7,8]{\\Miguel Aguilera}
\author[1,7,8]{Christopher Buckley}

\affil[1]{VERSES Research Labs, Los Angeles, California, USA}
\affil[2]{Department of Collective Behaviour, Max Planck Institute of \protect\\Animal Behavior, 78464 Konstanz, Germany}
\affil[3]{Department of Biology and Centre for the Advanced Study\protect\\of Collective Behaviour, University of Konstanz, 78464 Konstanz, Germany}
\affil[4]{Network Science Institute, Northeastern University, Boston, Massachusetts, USA}
\affil[5]{Laboratory for the Modeling of Biological and Socio-Technical Systems,\protect\\Northeastern University, Boston, Massachusetts, USA}
\affil[6]{Birkbeck Department of Film, Media and Cultural Studies, London, UK}
\affil[7]{Sussex AI Group, Department of Informatics, University of Sussex, Brighton, UK}
\affil[8]{Sackler Centre for Consciousness Science, University of Sussex, Brighton, UK}

\begin{document}
\maketitle
\pagenumbering{arabic}

\begin{abstract}
An open question in the study of emergent behaviour in multi-agent Bayesian systems is the relationship, if any, between individual and collective inference. In this paper we explore the correspondence between generative models that exist at two distinct scales, using spin glass models as a sandbox system to investigate this question. We show that the collective dynamics of a specific type of active inference agent is equivalent to sampling from the stationary distribution of a spin glass system. A collective of specifically-designed active inference agents can thus be described as implementing a form of sampling-based inference (namely, from a Boltzmann machine) at the higher level. However, this equivalence is very fragile, breaking upon simple modifications to the generative models of the individual agents or the nature of their interactions. We discuss the implications of this correspondence and its fragility for the study of multiscale systems composed of Bayesian agents.
\end{abstract}

\section{Introduction}
Emergent phenomena in multi-agent systems are central to the study of self-organizing, complex systems, yet the relationship between individual properties and group-level phenomena remains opaque and lacks formal explanation. One principled approach to understanding such phenomena is offered by the Free Energy Principle and so-called `multiscale active inference' \cite{friston2019free, ramstead2018answering}. Proponents of this multiscale approach propose that groups of individually-Bayesian agents necessarily entail an emergent, higher-order Bayesian agent --- in other words, systems are `agents all the way down' by definition \cite{kirchoff2018,palacios2020markov, ramstead2018answering, ramstead2019variational, hesp2019multi}. However, to date, there has been little theoretical or modeling work aimed at investigating whether this proposition is true in general or even demonstrating an existence proof in a specific system.

In this work we investigate this proposal by building a network of active inference agents that collectively implement a spin glass system. Spin glasses are a well-studied model class with a long history in statistical physics and equilibrium thermodynamics \cite{glauber1963time,brush1967history}. In the context of machine learning and computational neuroscience, spin glass systems can be tied to models of Bayesian inference and associative memory, particularly in the form of Hopfield networks and undirected graphical models like Boltzmann machines \cite{welling2003approximate, hopfield1982neural}. Boltzmann machines are a canonical example of an energy-based model in machine learning --- they are defined by a global energy function and are analytically equivalent to a particular sort of spin glass model. The Boltzmann machine can straightforwardly be shown to be an inferential model by conditioning on the states of particular spins and sampling from the posterior distribution over the remaining spins' states \cite{hinton1983optimal, hinton1986learning}. In doing so, Boltzmann machines and spin glass systems can be described as performing Bayesian inference about the latent causes (spin configurations) of the `data' (conditioned spins).

In this paper, we set out to investigate whether an inference machine (in our case, a Boltzmann machine) that exists at a `higher-level', can be hierarchically decomposed into an ensemble of agents collectively performing active inference at a `lower level.' We show a simple but rigorous equivalence between collective active inference and spin glass systems, providing the first steps for future quantitative study into the relationship between individual and collective generative models. We show that a group of active inference agents, equipped with a simple but very specific generative model, collectively sample from the stationary distribution of a spin glass system at the higher scale. This can be connected to a particular form of sampling known as Glauber dynamics \cite{glauber1963time, walter2015introduction}. When we further condition on the states of particular agents, then the system can be interpreted as collectively performing a form of sampling-based posterior inference over the configurations of the unconditioned agents, i.e., Boltzmann machine inference.

This paper is structured as follows: first, we specify the generative model that each spin site in a multi-agent spin glass system is equipped with, noting that the single agents are constructed explicitly such that their interactions at the lower-level lead to a spin glass system at the higher level. Then, we establish the equivalence between this multi-agent active inference process and Glauber dynamics, a scheme that samples from the stationary distribution of a spin glass system. We then generalize this result to sampling-based inference in Boltzmann machines by relaxing the homogeneous parameterization of each agent's generative models. We draw exact equivalences between the precisions of each agent's generative model and the parameters of a higher-level Boltzmann machine. We conclude by noting the fragility of the equivalence between multi-agent active inference and sampling from a collective spin glass system, and discuss the implications of our results for the study of emergent Bayesian inference in multiscale systems.

\section{Generative model for a single spin}

We begin by constructing a generative model for a single Bayesian agent, which we imagine as a single spin in a collective of similar agents. From the perspective of this single `focal agent' or spin, this generative model describes the agent's internal model of how the local environment generates its sensory data. Throughout this section we will take the perspective of this single focal agent, keeping in mind that any given agent is embedded in a larger system of other agents.

\subsection{States and observations}

We begin by specifying the state-space of observations and hidden states that characterize the focal agent's `world.' The focal agent's observations or sensations are comprised of a collection of binary spin states $\tilde{\sigma} = \{\sigma_j : j \in M\}$ where $\sigma_j = \pm1$ and $M$ is the set of other spins that the focal agent has direct access to. In other words, the agent directly observes the spin states of neighbouring agents (but not its own).

The focal agent assumes these observed spin states $\tilde{\sigma}$ all depend on a single, binary latent variable $z$ --- the `hidden spin state', which could also be interpreted as a coarse-grained `average spin' of its neighbours. Having specified observations $\tilde{\sigma}$ and latent states $z$, the full generative model can be written as a joint distribution over observations and the hidden spin state, $P(\tilde{\sigma}, z)$. This in turn factorizes into a set of distributions that describe the likelihood over observations, given the latent state $P(\tilde{\sigma}|z)$, and prior beliefs about the latent state $P(z)$:

\begin{align}
    P(\tilde{\sigma}, z) = P(\tilde{\sigma}|z)P(z)  \notag
\end{align}

We parameterize the likelihood and prior distributions as Bernoulli distributions (expressed in a convenient exponential form):

\begin{minipage}{0.45\textwidth}
  \begin{align}
      P(\tilde{\sigma}|z;\gamma) &= \prod_{j \in M} \frac{\exp(\gamma \sigma_j z)}{2 \cosh (\gamma z)} \notag \\
      &\text{Likelihood} \notag
    \end{align}
\end{minipage}
\begin{minipage}{0.45\textwidth}
    \begin{align}
      P(z; \zeta) &= \frac{\exp(\zeta z)}{2 \cosh (\zeta z)} \notag \\
      &\text{Prior} \notag
    \end{align}
\end{minipage}\vspace{0.35cm}

The likelihood factorizes into a product of independent Bernoulli distributions over each neighbouring spin $\sigma_j$. The full likelihood is parameterized with a single sensory precision parameter $\gamma$ whose magnitude captures the focal agent's assumption about how reliably neighbouring spin states $\sigma_j$ indicate the identity of the latent state $z$. A positive $\gamma$ indicates that $\sigma_j$ lends evidence to $z$ being aligned with $\sigma_j$, whereas a negative $\gamma$ means that $\sigma_j$ lends evidence to $z$ being opposite to $\sigma_j$. 

The prior over $z$ is also a Bernoulli distribution, parameterized by a precision $\zeta$ that acts as a `bias' in the focal agent's prior belief about the value of $z$. When $\zeta > 0$, the focal agent believes the `UP' ($z = +1$) state is more likely \textit{a priori}, whereas $\zeta < 0$ indicates that the agent believes that $z = -1$ is more likely, with the magnitude of $\zeta$ reflecting the strength or confidence of this prior belief.

\begin{figure}[t]
    \includegraphics[width=\textwidth]{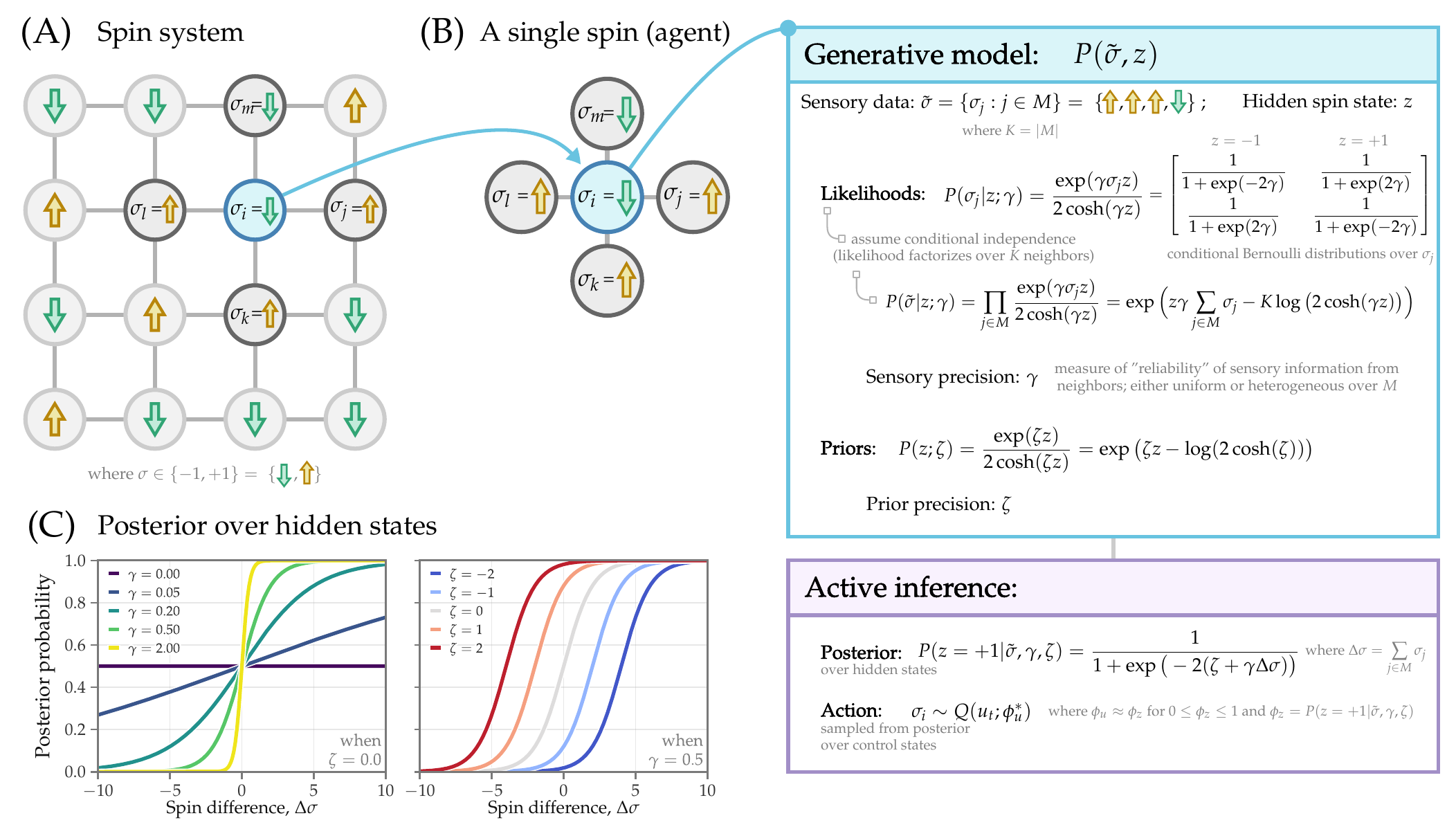}
	\caption{\textbf{Schematic illustration of individual and collective dynamics.} \textbf{(A)} Example of a system of 16 spin sites connected via a 2-D lattice, each in a state of $\sigma \in \{-1,+1\}$ (green down-arrow or yellow up-arrow above), with a focal agent and its spin, $\sigma_i = -1$, highlighted in blue. \textbf{(B)} Generative model of a single spin. \textbf{(C)} The posterior belief over $z = +1$ as a function of the spin difference $\Delta \sigma$. Left: The steepness of the function is tuned by $\gamma$ ($\zeta=0.0$ shown). Right: The horizontal shift depends on $\zeta$ ($\gamma=0.5$ shown).}
	\label{fig:schematic}
\end{figure}

\subsection{Bayesian inference of hidden states}

Having specified a generative model, we now consider (from the perspective of a focal agent) the problem of estimating $z$, given observations $\tilde{\sigma} = \{\sigma_j : j \in M\}$ and generative model parameters $\gamma, \zeta$. This is the problem of Bayesian inference, specifically the calculation of the posterior distribution over $z$. The conjugate-exponential form of the generative model means that the Bayesian posterior can be calculated exactly, and has a Bernoulli form that depends on $\tilde{\sigma}$ and $z$:

\begin{align}
\label{eq:full_posterior_formula}
    P(z | \tilde{\sigma}; \gamma, \zeta) = \frac{P(\tilde{\sigma}, z; \gamma, \zeta)}{P(\tilde{\sigma}; \gamma, \zeta)} = \frac{\exp\left( z (\zeta + \gamma \sum_{j} \sigma_j)\right)}{2\cosh\left(\zeta + \gamma \sum_{j} \sigma_j \right)}
\end{align}

If we fix the hidden state $z$ to a particular value (e.g. $z = +1$), then we arrive at a simple expression for the posterior probability that the hidden spin state is in the `UP' state, given the observations and the generative model parameters $\gamma, \zeta$. The posterior belief expressed as the sum of sensory input $\sum_j \sigma_j$ assumes a logistic or sigmoid form. Hereafter we refer to the sum of observed spin states as the `spin difference' $\Delta \sigma = \sum_{j} \sigma_j$, since this sum is equivalent to the difference in the number of positive ($\sigma_j = +1$) and negative ($\sigma_j = -1$) spins. Intuitively, the steepness and horizontal shift of this logistic function are determined by the likelihood and prior precisions:

\begin{align}
    P(z = +1|\tilde{\sigma}, \gamma, \zeta) = \frac{1}{1 + \exp\left(-2(\zeta + \gamma \Delta \sigma)\right)} \label{eq:prob_posterior_ON}
\end{align}

Figure \ref{fig:schematic}C shows the effect of varying the likelihood and prior precisions on the posterior belief over $z$ as a function of $\Delta \sigma$. We can also express the posterior as a Bernoulli distribution using the more common form, with parameter $\phi_z$:
\begin{align}
    P(z|\tilde{\sigma}, \gamma, \zeta ; \phi_z) &= (1 - \phi_z)^{1 - \frac{z + 1}{2}} \phi_z ^{\frac{z + 1}{2}} \notag \\
    \phi_z &= \frac{1}{1 + \exp(-2(\zeta + \gamma \Delta \sigma))}
\end{align}

We now have a simple update rule that expresses how a focal agent updates its beliefs about $z$ in response to observed spins $\tilde{\sigma}$. This sigmoid belief update has a clear, intuitive relationship to the parameters of the focal agent's generative model, with $\gamma$ encoding the sensitivity of the belief to small changes in $\Delta \sigma$ and $\zeta$ encoding a `bias' that skews the belief towards either $-1$ or $+1$. In the next section, we connect the generation of spins themselves to an active inference process, that leverages the Bayesian estimation problem of the current section to determine a focal agent's inference of its own spin state.

\subsection{Active inference of spin states}
Having addressed the issue of Bayesian inference or state estimation, we can now specify a mechanism by which agents generate their own spin states. These generated spin states will then serve as observations for the neighbours to whom the focal agent is connected. This turns into a problem of belief-guided action selection or decision-making. To enable agents to sample spin states as a function of their beliefs, we supplement each agent's current generative model with an extra random variable that corresponds to \textit{control states}, and some forward model of how those control states determine observations. We use \textit{active inference} to optimize a posterior belief over these control states \cite{friston2009reinforcement, friston2015active, friston2017active}; an agent can then act to change its spin state by sampling from this posterior. By equipping each agent with a particular type of forward model of how its actions impact observations, we can formally tie the collective dynamics of active inference agents with this generative model to a sampling scheme from a spin glass model. Appendix \ref{sec:appendix_B} walks through the steps needed to add a representation of control states into the generative model introduced in the previous section, and perform active inference with respect to this augmented generative model.

Active inference agents entertain posterior beliefs not only about the hidden states of the world, but also about how their own actions affect the world. Posterior beliefs about actions are denoted $Q(u ; \phi_u)$, where $u$ is a random variable corresponding to actions and $\phi_u$ are the parameters of this belief. As opposed to the analytic posterior over hidden states $z$, $Q(u;\phi_u)$ is an approximate posterior, optimized using variational Bayesian inference \cite{blei2017variational}. In our focal agent's simple action model, control states have the same support as hidden states, i.e. $u = \pm 1$. The value of $u$ represents a possible spin action to take (`UP' vs. `DOWN'). We parameterize $Q(u;\phi_u)$ as a Bernoulli distribution with parameter $\phi_u$, which itself encodes the probability of taking the `UP' ($+1$) action:

\begin{align}
    Q(u_t;\phi_u) = (1-\phi_z)^{1 - \frac{u_t + 1}{2}} \phi_z^{ \frac{u_t + 1}{2}} 
\end{align}

When we equip our spin glass agents with a particular (predictive) generative model, we can show that the approximate posterior over control states simplifies to the state posterior (see Appendix \ref{sec:appendix_B} for details), and an agent can generate a spin state by simply sampling from the posterior over actions:

\begin{minipage}{0.45\textwidth}
  \begin{align}
       Q(u;\phi_u) &\approx P(z|\tilde{\sigma}, \gamma, \zeta ; \phi_z) \notag \\
        \phi_u &\approx \phi_z: 0 \leq \phi_z \leq 1 \notag
    \end{align}
\end{minipage}
\begin{minipage}{0.5\textwidth}
    \begin{align}
    \label{eq:action_sampling_rule}
      \sigma &\sim Q(u;\phi_u) \notag \\
    &\sim Q(z;\phi_z) \triangleq P(z|\tilde{\sigma};\gamma, \zeta)
    \end{align}
\end{minipage}
\vspace{0.35cm}

We now have an active inference agent that 1) calculates a posterior belief $P(z|\tilde{\sigma}, \gamma, \zeta; \phi_z)$ about the latent state $z$ in response to the observed spins of other agents and 2) generates a spin of its own by sampling from this belief, which ends up being identical to the posterior over actions. Intuitively, each agent just broadcasts its beliefs about the latent cause of its social observations, by sampling from its posterior over this hidden (average) state. Another way of looking at this is that each agent emits actions that maximize the accuracy of its beliefs (i.e., minimize variational free energy), under the prior assumption it is the author of its sensations, which, implicitly, are shared with other agents. Note that the choice to sample from the posterior over actions (as opposed to e.g.~taking the maximum) renders this action-selection scheme a form of probability matching \cite{shanks2002re, perez2011collective}. 

\subsection{Completing the loop}
Given this sampling scheme for generating actions, we can simulate a collective active inference process by equating the actions of one agent to the observations of another. Specifically, each focal agent's spin action becomes an observation ($\sigma_j$ for some $j$) for all the other agents that it (the focal agent) is connected to. Next, we will show how the dynamics of multi-agent active inference is analogous to a particular algorithm for sampling from the stationary distribution of a spin glass model, known as Glauber dynamics \cite{glauber1963time}. We then examine the fragility of this equivalence by exploring a number of simple modifications that break it.

\section{Equivalence to Glauber dynamics}

Spin glass models are formally described in terms of a global energy function over states of the system. The global energy is related to the stationary probability distribution of the system through a Gibbs law or Boltzmann distribution:

\begin{align}
    \label{eq:gibbs_measure}
    p^{*}(\tilde{\sigma}) &= \frac{1}{Z}\exp(-\beta E(\tilde{\sigma}))
\end{align}
where the stationary density $p^{*}$ and energy function $E$ are defined over spin configurations, where a configuration is a particular setting of each of the $N$ spins that comprise the system: $\tilde{\sigma} = \{\sigma_i\}_{i = 1}^{N}: \sigma_i \pm 1$. The partition function $Z$ is a normalizing constant that ensures $p^{*}(\tilde{\sigma})$ integrates to $1.0$, and $\beta$ plays the role of an inverse temperature that can be used to arbitrarily rescale the Gibbs measure. In the case of the Ising model, this energy function is a Hamiltonian that can be expressed as a sum of pairwise interaction terms and an external drive or bias (often analogized to an external magnetic field):

\begin{align}
    E(\tilde{\sigma}) &= -\sum_{\langle i, j \rangle} \sigma_i J_{ij} \sigma_j - \sum_{i} h_i \sigma_i  \label{eq:ising_energy_func}
\end{align}
where $J_{ij}$ specifies a (symmetric) coupling between spin sites $i$ and $j$ and $h_i$ specifies an external forcing or bias term for site $i$. The bracket notation $\langle i, j \rangle$ denotes a sum over pairs. In numerical studies of the Ising model, one is typically interested in generating samples from this stationary density. One scheme for doing so is known as Glauber dynamics, where each spin $\sigma_i$ of the system is updated using the following stochastic update rule:

\begin{align}
\label{eq:glauber_sampling_rule}
    \sigma_i &\sim P(\sigma_i = (-1, +1)) \notag \\
    P(\sigma_i = +1) &= \frac{1}{1 + \exp(-\beta \Delta_i E)} \notag \\
    \Delta_i E &= E_{\sigma_i = DOWN} - E_{\sigma_i = UP} 
    = 2\left(\sum_{j \in M_i} J_{ij}\sigma_j + h_i\right)
\end{align}
where $\Delta_i E$ represents the difference in the energy between configurations where $\sigma_i = -1$ and those where $\sigma_i = +1$. In other words, the probability of unit $i$ flipping to $+1$ is proportional to the degree to which the global energy $E$ would decrease as a result of the flip. If units are updated sequentially (also known as `asynchronous updates'), then given sufficient time Glauber dynamics are guaranteed to sample from the stationary density in Equation \eqref{eq:gibbs_measure} \cite{glauber1963time}.

The probability of agent $i$ taking action $\sigma_i = +1$ is given by the action sampling rule in Equation \eqref{eq:action_sampling_rule} of the previous section. We can thus write the probability of taking a particular action in terms of the posterior over latent states $z$, by plugging in the posterior belief over $z = +1$ (given in Equation \eqref{eq:prob_posterior_ON}) into Equation \eqref{eq:action_sampling_rule}:

\begin{align}
\label{eq:sampling_spin_posterior_states}
    P(\sigma_i = +1) &= \frac{1}{1 + \exp(-2(\zeta_i + \gamma_i \Delta_i\sigma))}
\end{align}
where we now index $\zeta, \gamma, \Delta\sigma$ by $i$ to indicate that these are the generative model parameters and observations of agent $i$. The identical forms shared by Equations \eqref{eq:glauber_sampling_rule} and \eqref{eq:sampling_spin_posterior_states} allow us to directly relate the parameters of individual generative models to the local energy difference $\Delta_i E$ and the global energy of the system.

\begin{align}
    \Delta_i E \propto J \sum_{j \in M_i} \sigma_j + h_i = \gamma \Delta_i \sigma + \zeta_i \notag
    \implies J = \gamma, \hspace{2mm} h_i = \zeta_i
\end{align}
where we assume all agents share the same likelihood precision $\gamma_i = \gamma^{*}: \forall i$, which is equivalent to forcing all couplings to be identical $J_{ij} = J: \forall i, j$.  Individual active inference agents in this multi-agent dynamic thus behave as if they are sampling spin states in order to minimize a global energy function defined over spin configurations, which in the case of spin glass systems like the Ising model, can be computed using local observations (the spins of one's neighbours) and model parameters $\gamma, \zeta$. Going the other direction, one can sample from the stationary distribution of spin glass system by simulating a collective of active inference agents with an appropriately parameterized generative model. 

However, the equivalence between Glauber sampling from the stationary distribution of an Ising model and collective active inference breaks down when agents update their actions in parallel or synchronously, rather than asynchronously. In particular, under parallel action updates, the system samples from a stationary distribution with a different energy function than that from Equation \eqref{eq:ising_energy_func}. See Appendix \ref{sec:appendix_C} for derivations on the relationship between the schedule of action updates (synchronous vs. asynchronous) and the resulting stationary density of the system.

\section{Equivalence to inference in Boltzmann machines}\label{sec:boltzmann_equivalence}

Connecting the collective dynamics of these specialized active inference agents to inference in Boltzmann machines is straightforward. We now equip each agent's generative model with a \textit{vector} of sensory precisions $\tilde{\gamma} = \{\gamma_j: j \in M\}$ that act as different reliabilities assigned to different neighbours' spin observations. The new factorized likelihood can be written:

\begin{align}
\label{eq:boltzmann_likelihood}
    P(\tilde{\sigma}|z; \tilde{\gamma}) = \prod_{j \in M} \frac{\exp (\gamma_j \sigma_j z)}{2 \cosh (\gamma_j z)}
\end{align}

We can then write the posterior over $z$ as a function of observations and generative model parameters $\{\tilde{\sigma}, \zeta, \tilde{\gamma} \}$. By fixing $z$ to the value $+1$, we obtain again a logistic expression for the posterior probability $P(z = +1|\tilde{\sigma}, \tilde{\gamma}, \zeta)$ that is nearly identical to the original Equation \eqref{eq:prob_posterior_ON}:

\begin{align}
\label{eq:boltzmann_posterior_z_fixed}
    P(z = +1 | \tilde{\sigma}, \tilde{\gamma}, \zeta) = \frac{1}{1 + \exp(-2(\zeta + \sum_{j \in M} \gamma_j \sigma_j))}
\end{align}

A Boltzmann machine is a special variant of a spin glass model, defined by a global energy that in turn defines the stationary probability assigned to each of the system's configurations. The Boltzmann energy $E_{B}$, as with the classical spin glass energy, is defined over configurations of the system's binary units (also known as nodes or neurons) $\tilde{x} = \{x_i\}_{i = 1}^{N}$.

In the context of inference, it is common to partition the system's units into `visible units' and `hidden units', $\tilde{x} = \{\tilde{v}, \tilde{h} \}$, with the following energy function: 

\begin{align}
    \label{eq:boltzmann_global_energy_partitioned}
    E_{B}(\tilde{v}, \tilde{h}) &= - \frac{1}{2}(\tilde{v}^{\top}\mathbf{W}_{vv}\tilde{v} + \tilde{h}^{\top}\mathbf{W}_{hh}\tilde{h} + \tilde{v}^{\top}\mathbf{W}_{vh}\tilde{h}) - \tilde{\theta}^{\top}_v\tilde{v} - \tilde{\theta}^{\top}_h\tilde{h}
\end{align}
where $\mathbf{W}_{vv}, \mathbf{W}_{hh}, \mathbf{W}_{vh}$ are weight matrices with symmetric couplings between units with no `self-edges' ($W_{ii} = 0: \forall i$) that mediate dependencies both within and between the two subsets of units $\tilde{v}, \tilde{h}$; and $\tilde{\theta}_v, \tilde{\theta}_h$ are vectors of unit-specific biases or thresholds.   The Bayesian inference interpretation of Boltzmann machines considers the conditional probability distribution over $\tilde{h}$, given some fixed values of $\tilde{v}$. The `clamping' of visible nodes to some data vector $\tilde{v} = \tilde{d}$ can simply be absorbed into the biases $\tilde{\theta}_v$, such that samples from the posterior $P(\tilde{h}|\tilde{v} = \tilde{d})$ are analogous to sampling from the joint distribution $P(\tilde{h}, \tilde{v})$ where the biases of the visible nodes are adjusted to reflect this clamping. Sampling from this model can be achieved with Glauber dynamics, since the model is a spin glass system with heterogeneous (but symmetric) couplings. We can therefore write the single unit `ON' probability as follows, now in terms of weights $\mathbf{W}$ and thresholds $\tilde{\theta}$:

\begin{align}
    P(x_i = +1) &= \frac{1}{1 + \exp (-\Delta_i E_{B})}
    = \frac{1}{1 + \exp (-\sum_{j} W_{ij} x_j - \theta_i)}
\end{align}
where the interaction term that comprises the local energy difference $\Delta_i E_{B}$ is equivalent to a dot-product between the $i\textsuperscript{th}$ row of $\mathbf{W}$ and vector of activities $\tilde{x}$. It is thus straightforward to relate the weights and biases of a Boltzmann machine to the sensory and prior precisions of each agent's generative model. In particular, the weight connecting unit $j$ to unit $i$ in a Boltzmann machine $W_{ij}$ is linearly related to the precision that agent $i$ associates to spin observations coming from agent $j$: $W_{ij} = 2\gamma_{(i, j)}$ where the subscript $(i, j)$ refers to agent $i$'s likelihood model over observations emitted by agent $j$. If agents $i$ and $j$ are not connected, then $W_{ij} = \gamma_{(i,j)} = 0$. The bias of the $i$\textsuperscript{th} unit is also straightforwardly related to agent $i$'s prior precision via $\theta_i = 2\zeta_i$.

We have seen how sampling from the posterior distribution of a Boltzmann machine with fixed visible nodes $\tilde{v}$ is equivalent to the collective dynamics of a specific multi-agent active inference scheme. We have thus shown a carefully-constructed system, in which a form of sampling-based Bayesian inference at one scale emerges from a process of collective active inference at a lower scale.

\section{Discussion}

Although the equivalences we have shown are exact, there are numerous assumptions that, when broken, violate the equivalence between the multi-agent dynamics defined at the lower level, and the higher-level sampling dynamics of the spin glass system.

The energy-based models we have studied (Ising models, Boltzmann machines) are all \textit{undirected} graphical models: this means that the global energy function is defined by symmetric interaction terms across pairs of spins: $J_{ij} = J_{ji}$\footnote{$\mathbf{W} = \mathbf{W}^{\top}$ for the Boltzmann machine, respectively}. In order to meet this requirement at the level of the individual agents, one must force the precisions that a pair of agents assign to one another to be identical: $\gamma_{(i, j)} = \gamma_{(j, i)}$. This constraint also underpins the equilibrium nature of classical spin glass systems where detailed balance conditions are met, i.e., the system is in thermodynamic equilibrium. In natural complex systems (especially biological ones), these detailed balance conditions are often broken, and the systems operate far from equilibrium \cite{kwon2011nonequilibrium, yan2013nonequilibrium,ma2014potential, millan2020explosive, friston2021stochastic, aguilera2021unifying, lynn2020broken, aguilera2022nonequilibrium}. This may manifest in the case of realistic multi-agent dynamics in the form of asymmetric beliefs about reliability of social signals that agents assign to one another \cite{albarracin2022epistemic}.

Another fragility of the multiscale equivalence is the structure of the individual generative model, which relies on very specific assumptions about how hidden states $z$ relate to observations. Without this generative model at the single-agent level, there is no equivalence between the collective active inference process and a spin glass model --- the model's dynamics could become more complex (and potentially analytically intractable), because the posterior update is not guaranteed to be a simple logistic function of the sum of neighbouring spins. This could be easily shown by changing the single agent's likelihood model to include a separate latent variable for each neighbour's spin state\footnote{This is analogous to the approach taken in \cite{albarracin2022epistemic}, where each agent had beliefs about the belief state of each of its neighbours}, or if the total likelihood over neighbouring spin observations did not factorize into a product of neighbour-specific likelihoods, but had some more complex statistical structure.

Finally, the convergence of Glauber dynamics to samples from the joint density over spin configurations depends on the temporal schedule of action updating; namely, spins have to be updated sequentially or asynchronously, rather than in parallel (see Appendix \ref{sec:appendix_C} for details) in order to guarantee sampling from the stationary distribution in Equation \eqref{eq:gibbs_measure}. If agents act in parallel, then the stationary distribution of spin states is different than that given by the classical spin glass Hamiltonian. In other words, depending on the relative timescales of collective action, agents either will or will not minimize the global energy function that their actions appear to local minimize (i.e. actions that minimize the local energy difference $\Delta_i E$).

\section{Conclusion}

In this work we demonstrate an exact equivalence between a collective active inference process and sampling from the stationary density of a spin glass system. Furthermore, we connect the system's collective dynamics to sampling-based inference in energy-based models like Boltzmann machines. In particular, when we constrain certain agents in the network to be `visible nodes' and fix their actions, then the whole system samples from the posterior distribution over spin configurations, given the actions of the `visible' agents. Despite these exact relationships, we also note the fragility of the equivalence, which relies on very particular assumptions. These include the symmetry of the precisions that pairs of agents assign to each other, the temporal scheduling of the action updates, and the specific generative model used by the agents. It remains to be seen, whether when these assumptions are broken,  an inferential or `agentive' interpretation still obtains at higher scales, and if so, whether the form of the `collective' generative model can be analytically related to the individual generative models as it was in the present, equilibrium case.

Our results have important implications for the overall agenda of multiscale active inference, and the quest to uncover the quantitative relationship between generative models operating at distinct scales in complex systems. In the system presented in the current work, we show that active inference agents may collectively achieve sampling-based inference at a distinct, higher level under particular conditions. Despite the apparent consistency at the two scales, our result actually conflicts with claims made in the multiscale active inference literature, that posits that systems hierarchically decompose into nested levels of active inference agents \cite{ramstead2018answering, kirchoff2018, ramstead2019variational, friston2019free, palacios2020markov} --- in other words, that systems are inherently active inference processes `all the way down.' Note that in our system, there are only active inference agents operating at the lower level --- the higher level is not an active inference agent, but is better described as a passive agent that performs hidden state-estimation or inference by sampling from a posterior belief over spin configurations. The agenda of the present work also resonates with ongoing research into the necessary and sufficient conditions for generic complex systems to be considered `agentive' or exhibit inferential capacities \cite{virgo2021interpreting, krakauer2020information, krafft2021bayesian}.

Our results suggest that multiscale inference interpretations of complex systems do not necessarily emerge in any system. We nevertheless hope that the simple equilibrium case we presented here may serve as a launching pad for future studies into whether inference interpretations can be rescued at the higher scale in cases when the fragile assumptions at the single-agent level are broken. 

\section*{Additional information}\label{sec:additional_info}

\paragraph{Acknowledgements}
The authors thank Alex Kiefer, Beren Millidge, Dalton Sakthivadivel, Magnus Koudahl, Dimitrije Markovic and Maxwell Ramstead for their feedback and comments throughout the writing of the manuscript.

\paragraph{Funding information}
C.H. is supported by the U.S. Office of Naval Research (N00014-19-1-2556). C.H., B.K., \& D.D. acknowledge the support of a grant from the John Templeton Foundation (61780). The opinions expressed in this publication are those of the author(s) and do not necessarily reflect the views of the John Templeton Foundation.

\printbibliography[title={References}]

\clearpage

\appendix
\setcounter{figure}{0}
\setcounter{table}{0}
\setcounter{equation}{0}
\renewcommand\thefigure{\thesection.\arabic{figure}}
\renewcommand\thetable{\thesection.\arabic{table}}
\renewcommand\theequation{\thesection .\arabic{equation}}
\begin{refsection}

\section*{Appendix: Spin glass systems as collective active inference}

\section{Bayesian inference for a single agent}\label{sec:appendix_A}

In this appendix we derive the exact Bayesian inference update for the posterior over the latent state $z$, taking the perspective of a single agent. We begin by rehearsing the component likelihood and prior distributions of the generative model in more detail.

\subsection{Likelihood}

The likelihood model relates the hidden state $z$ to the observed spin state of a particular neighbour $\sigma_j$ as an exponential distribution parameterized by a sensory precision parameter $\gamma$:
\begin{align}
    P(\sigma_j|z; \gamma) = \frac{\exp (\gamma \sigma_j z)}{2 \cosh (\gamma z)} \label{eq:appA_single_neighbour_likelihood}
\end{align}

The sign and magnitude of $\gamma$ determines the nature of the expected mapping between hidden states $z$ and neighbouring spin observations $\sigma_j$. For $\gamma > 0$, then the observed spin is expected to reflect the latent state $z$, and with $\gamma < 0$, then the observed spin is expected to be opposite to the latent state $z$. The magnitude of $\gamma$ then determines how deterministic this mapping is.

Equation \eqref{eq:appA_single_neighbour_likelihood} can alternatively be seen as a collection of two conditional Bernoulli distributions over $\sigma_j$, one for each setting of $z$. This can be visualized as a symmetric matrix mapping from the two settings of $z$ (the columns, corresponding to $z = -1, +1$) to the values of $\sigma_j$ (the rows $\sigma_j = -1, +1$):
\begin{align}
    P(\sigma_j | z ; \gamma) &= \begin{bmatrix} \dfrac{1}{1 + \exp(-2\gamma)} & \dfrac{1}{1 + \exp(2\gamma)} \\[0.19cm] \dfrac{1}{1 + \exp(2\gamma)} & \dfrac{1}{1 + \exp(-2\gamma)} \end{bmatrix}
\end{align}
where this mapping approaches the identity matrix as $\gamma \to \infty$.

Now we can move onto writing down the likelihood over the observed spins of multiple neighbours: $\tilde{\sigma} = \{\sigma_j : j \in M\}$ where $M$ denotes the set of the focal agent's neighbours. We build in a \textit{conditional independence} assumption into the focal agent's generative model, whereby the full likelihood model over all observed spins factorizes across the agent's neighbours. This means we can write the likelihood as a product of the single-neighbour likelihoods shown in Equation \eqref{eq:appA_single_neighbour_likelihood}:
\begin{align}
    P(\tilde{\sigma}|z; \gamma) 
    &= \prod_{j \in M} \frac{\exp (\gamma \sigma_j z)}{2 \cosh (\gamma z)} \notag \\
    &= \exp\left(z \gamma \sum_{j \in M} 
    \sigma_j  - K \log (2 \cosh(\gamma z))\right) \label{eq:appA_full_likelihood}
\end{align}
where $K$ is the number of the focal agent's neighbours (i.e. the size of the set $M$).
We can easily generalize this likelihood to heterogeneous precisions by instead parameterizing it with a precision vector $\tilde{\gamma} = \{\gamma_j : j \in M\}$ that assigns a different precision to observations coming from each of the focal agent's neighbours:
\begin{align}
    P(\tilde{\sigma}|z;\tilde{\gamma}) &= \prod_{j \in M} \frac{\exp(\gamma_j \sigma_j z)}{2 \cosh(\gamma_j z)} \notag \\
    &= \exp\left(z \sum_{j \in M}\gamma_j \sigma_j - \sum_{j \in M} \log (2 \cosh(\gamma_j z))\right)
\end{align}

\subsection{Prior}
We parameterize the focal agent's prior beliefs about the latent spin state $z$ as a simple Bernoulli distribution, and similarly to the likelihood model, we will express it as an exponential function parameterized by a `prior precision' parameter $\zeta$:
\begin{align}
    P(z;\zeta) &= \frac{\exp(\zeta z)}{2 \cosh (\zeta z)}\notag = \exp(\zeta z - \log( 2 \cosh (\zeta ))) \label{eq:appA_exponential_bernoulli_prior}
\end{align}

As with the sensory precision $\gamma$, the prior precision also scales the strength of the focal agent's prior belief that the spin state $z$ is $+1$.\footnote{Note that $\cosh(\zeta z)$ can be re-written $\cosh(\zeta)$ when $z \in \{-1, +1 \}$ due to the symmetry of the hyperbolic cosine function around the origin.}

\subsection{Bayesian inference of hidden states}

Now we ask the question: how would a focal agent (i.e., the agent that occupies a single lattice site) optimally compute a belief over $z$, that is most consistent with a set of observed spins $\tilde{\sigma}$? This is a problem of Bayesian inference, which can be expressed as calculation of the posterior distribution over $z$ via Bayes Rule:

\begin{align}
    P(z | \tilde{\sigma}; \gamma, \zeta) = \frac{P(\tilde{\sigma}, z; \gamma, \zeta)}{P(\tilde{\sigma}; \gamma, \zeta)}
\end{align}

Since we are dealing with a conjugate exponential model \footnote{The Bernoulli prior is conjugate to the likelihood model, which can also be described of as a set of conditional Bernoulli distributions.}, we can derive an analytic form for the posterior:
$P(z | \tilde{\sigma}, \gamma, \zeta)$:

\begin{align}
    P(z|\tilde{\sigma}; \gamma, \zeta)
    = \frac{\exp\left( z (\zeta + \gamma \sum_{j} \sigma_j)\right)}{2\cosh\left(\zeta + \gamma \sum_{j} \sigma_j \right)} \label{eq:appA_posterior_exponential_form}
\end{align}
where the sum over neighbouring spins $j$ only includes the neighbours of the focal agent, i.e, $\sum_{j \in M} \sigma_j$.
If we fix the hidden state $z$ to a particular value (e.g. $z = +1$), then we arrive at a simple expression for the posterior probability that the hidden spin state is in the `UP' state, given the observations and the generative model parameters $\gamma, \zeta$. This probability reduces to a logistic or sigmoid function of sensory input, which is simply the sum of neighbouring spin values $\Delta \sigma = \sum_{j} \sigma_j$. This can also be seen as the `spin difference', or the number of neighbouring spins that are in the `UP' position, minus those that are in the `DOWN' position. The steepness and horizontal shift of this logistic function are intuitively given by likelihood and prior precisions, respectively:

\begin{align}
    P(z = +1|\tilde{\sigma}, \gamma, \zeta) &= \frac{\exp(\zeta + \gamma \Delta \sigma)}{\exp(\zeta + \gamma \Delta \sigma) + \exp(- (\zeta + \gamma \Delta \sigma))} \notag \\
    &= \left(1 + \frac{\exp(-(\zeta + \gamma \Delta \sigma))}{\exp(\zeta + \gamma \Delta \sigma)}\right)^{-1} \notag \\
    &= \frac{1}{1 + \exp(-2(\zeta + \gamma \Delta \sigma))} \label{eq:appA_prob_posterior_ON}
\end{align}

The denominator in the first line of \eqref{eq:appA_prob_posterior_ON} follows from the identity $\cosh(x) = \frac{1}{2}(\exp(x) + \exp(-x))$.

\section{\label{sec:appendix_B}Active inference derivations}

In this section we provide the additional derivations needed to equip each agent with the ability to infer a posterior over control states and sample from this posterior to generate actions. This achieved through the framework of \textit{active inference}.

Active inference casts the selection of control states or actions as an inference problem, whereby actions $u$ are sampled or drawn from posterior belief about controllable hidden states. The posterior over actions is computed as the softmax transform of a quantity called the \textit{expected free energy} \cite{friston2015active}. This is the critical objective function for actions that enables active inference agents to plan actions into the future, since the expected free energy scores the utility of the anticipated consequences of actions.

\subsection{Predictive generative model}

We begin by writing a so-called `predictive' generative model that crucially includes probability distributions over the agent's own control states $u \in \{-1, +1\}$ and how those control states relate to future (anticipated) observations. In other words, we consider a generative model over two timesteps: the current timestep $t$ and one timestep in the future, $t+1$. This will endow our agents with a shallow form of `planning', where they choose actions in order to maximize some (pseudo-) reward function defined with respect to expected outcomes. This can be expressed as follows:

\begin{align}
    P(\tilde{\sigma}_t, z_{t}, u_{t}, \mathcal{O}_{t+1}; \gamma, \zeta) = \tilde{P}( \mathcal{O}_{t+1}|z_t, u_t, \tilde{\sigma}_{t})P(\tilde{\sigma}_t, z_t, u_t; \gamma, \zeta) 
\end{align}
where the generative model at the second timestep $\tilde{P}( \mathcal{O}_{t+1}|z_t, u_t, \tilde{\sigma}_{t})$ we hereafter refer to as the `predictive' generative model, defined over a single future timestep.

Active inference consists in sampling a belief from the posterior distribution over control states $u$ --- this sampled control state or action is then fixed to be the spin state of the agent under consideration. Thus the action of one agent is fed in as the observations for those spin sites that it's connected to. In order to imbue active inference agents with a sense of goal-directedness or purpose, we encode a prior distribution over actions $P(u)$ that is proportional to the negative of the expected free energy, via the softmax relationship:

\begin{align}
    P(u) = \frac{\exp(-\mathbf{G}(u))}{ \sum_u \exp(-\mathbf{G}(u))}
\end{align}

Crucially, the expected free energy of an action $\mathbf{G}$ is a function of outcomes \textit{expected} under a particular control state $u$,  where beliefs about future outcomes are `biased` by prior beliefs about encountering particular states of affairs. In order to optimistically `bend' these future beliefs towards certain outcomes, and thus make some actions more probable than others, we supplement the predictive generative model $\tilde{P}$ with a binary `optimality' variable $\mathcal{O} \pm 1$ that the agent has an inherent belief that it will observe. This is encoded via a `goal prior' or preference vector, which is a Bernoulli distribution over seeing a particular value of $\mathcal{O}$ with some precision parameter $\omega$:

\begin{align}
    \tilde{P}(\mathcal{O}_{t+1}; \omega) &= \frac{\exp(\omega \mathcal{O})}{2 \cosh (\omega \mathcal{O})}
\end{align}

Hereafter we assume an infinitely high precision, i.e. $\omega \to \infty$. This renders the preference an `all-or-nothing' distribution over observing the optimality variable being in the `positive' state $\mathcal{O} = +1$:

\begin{align}
     &=\begin{bmatrix} \tilde{P}(\mathcal{O}_{t+1} = -1) \\ \tilde{P}(\mathcal{O}_{t+1} = +1)\end{bmatrix} = \begin{bmatrix} 0.0 \\ 1.0 \end{bmatrix}
\end{align}

To allow an agent the ability to predict the relationship between their actions and expected observations, it's important to include an additional likelihood distribution, what we might call the `forward model' of actions $P(\mathcal{O}_{t+1}|z_t, u_t; \xi)$. This additional likelihood encodes the focal agent's assumptions about the relationship between hidden states, actions, and the (expected) optimality variable. By encoding a deterministic conditional dependence relationship into this likelihood, we motivate the agent (via the expected free energy) to honestly signal its own estimate of the hidden state via its spin action $u$. To achieve this, we explicitly design this likelihood to have the following structure, wherein the optimality variable is only expected to take its `desired value' of $\mathcal{O} = +1$ when $z = u$. This can be written as a set of conditional Bernoulli distributions over $\mathcal{O}$, and each of which jointly depends on $z$ and $u$ and is parameterized by a (infinitely high) precision $\xi$:

\begin{align}
    P(\mathcal{O}_{t+1}  | z_{t} , u_{t} ; \xi) &= \frac{\exp(\xi \mathcal{O}_{t+1}  z_t u_t)}{2\cosh (\xi z_t u_t)}
\end{align}

When we assume $\xi \to \infty$, then we arrive at a form for this likelihood which can be alternatively expressed as a set of Bernoulli distributions that conjunctively depend on $z$ and $u$, and can be visualized as follows:

\begin{align}
    P(\mathcal{O}_{t+1}  | z_t, u_t = -1) &= \begin{bmatrix} 0 & 1 \\ 1 & 0  \end{bmatrix} \notag \\
     P(\mathcal{O}_{t+1}  | z_t, u_t = +1) &= \begin{bmatrix} 1 & 0 \\ 0 & 1 \end{bmatrix} \notag \\
\end{align}
where the columns of the matrices above correspond to settings of $z \in \{-1, +1\}$. Therefore, the agent only expects to see $\mathcal{O} = +1$ (the desired outcome) when the value of the hidden state and the value of the control variable are equal, i.e. $z = u$; otherwise $\mathcal{O} = -1$ is expected. For the purposes of the present study, we assume both the optimality prior $\tilde{P}(\mathcal{O};\omega)$ and the optimality variable likelihood $P(\mathcal{O}|z,u; \xi)$ are parameterized by infinitely high precisions $\omega = \xi = \infty$, and hereafter will exclude them when referring to these distributions for notational convenience.

Having specified these addition priors and likelihoods, we can write down the new (predictive) generative model as follows:

\begin{align}
    \tilde{P}(\mathcal{O}_{t+1}, u_t , z_t) = P(\mathcal{O}_{t+1} | z_t, u_t) P(u_t)\tilde{P}(\mathcal{O}_{t+1})P(z_t)
\end{align}

\subsection{Active inference}

Under active inference, both state estimation and action are consequences of the optimization of an approximate posterior belief over hidden states and actions $Q(z, u;\phi)$. This approximate posterior is optimized in order to minimize a variational free energy (or alternatively maximize an evidence lower bound). This is the critical concept for a type of approximate Bayesian inference known as variational Bayesian inference \cite{blei2017variational}. This can be described as finding the optimal set of variational parameters $\phi$ that minimizes the following quantity:

\begin{align}
    \phi^{*} &= \underset{\phi}{\arg \min} \, \, \mathcal{F}  \notag \\ &= \mathbb{E}_Q[\ln Q(z_t, u_t;\phi) - \ln \tilde{P}(\tilde{\sigma}_t, z_t, u_t, \mathcal{O}_{t+1}; \gamma, \zeta)]
\end{align}

In practice, because of the factorization of the generative model into a generative model of the current and future timesteps, we can split state-estimation and action inference into two separate optimization procedures. To do this we also need to factorize the posterior as follows:

\begin{align}
    Q(z, u;\phi) = Q(z;\phi_z)Q(u;\phi_u)
\end{align}
where we have also separated the variational parameters $\phi = \{\phi_z, \phi_u\}$ into those that parameterize the belief about hidden states $\phi_z$, and those that parameterize the belief about actions $\phi_u$.

When considering state-estimation (i.e. optimization of $Q(z_t;\phi_z)$), we only have to consider the generative model of the current timestep $P(\tilde{\sigma}_t, z_t; \gamma, \zeta)$. The optimal posterior parameters $\phi^{*}_z$ are found as the minimum of the variational free energy, re-written using only those terms that depend on $\phi_z$:

\begin{align}
   \phi^{*}_z &= \underset{\phi_z}{\arg \min} \,\, \mathcal{F}(\phi_z) \notag \\ \mathcal{F}(\phi_z) &=  \mathbb{E}_{Q(z_t ; \phi_z)}[\ln Q(z_t;\phi_z) - \ln P(\tilde{\sigma}_t, z_t; \gamma, \zeta)] \label{eq:appB_vfe_problem_qz}
\end{align}

To solve this, we also need to decide on a parameterization of the approximate posterior over hidden states $z_t$. We parameterize $Q(z_t;\phi_z)$ as a Bernoulli distribution with parameter $\phi_z$, that can be interpreted as the posterior probability that $z_t$ is in the `UP' ($+1$) state:

\begin{align}
    Q(z_t;\phi_z) = (1-\phi_z)^{1 - \frac{z_t + 1}{2}} \phi_z^{ \frac{z_t + 1}{2}} 
\end{align}

By minimizing the variational free energy with respect to $\phi_z$, we can obtain an expression for the optimal posterior $Q(z;\phi^{*}_z)$ that sits at the variational free energy minimum. Due to the exponential and conjugate form of the generative model, $Q(z_t;\phi_z)$ is the exact posterior and thus variational inference reduces to exact Bayesian inference. This means we can simply re-use the posterior update equation of Equation \eqref{eq:appA_prob_posterior_ON} to yield an analytic expression for $\phi^{*}_z$:

\begin{align}
    \phi^{*}_z = \frac{1}{1 + \exp\left(-2(\zeta + \gamma \Delta \sigma)\right)}
\end{align}

When considering inference of actions, we now consider the generative model of the future timestep, which crucially depends on the current control state $u_t$ and the optimality variable $\mathcal{O}_{t+1}$. We can then write the variational problem as finding the setting of $\phi_u$ that minimizes the variational free energy, now re-written in terms of its dependence on $\phi_u$:

\begin{align}
   \phi^{*}_u &= \underset{\phi_u}{\arg \min} \,\, \mathcal{F}(\phi_u) \notag \\ \mathcal{F}(\phi_u) &=  \mathbb{E}_{Q(u_t ; \phi_u)}[\ln Q(u_t;\phi_u) - \ln \tilde{P}(\mathcal{O}_{t+1}, u_t, z_t)] \label{eq:appB_vfe_problem_qu}
\end{align}

As we did for the posterior over hidden states, we need to decide on a parameterization for the posterior over actions $Q(u_t ; \phi_u)$; we also parameterize this as a Bernoulli distribution with parameter $\phi_u$ that represents the probability of taking the `UP' ($+1$) action:

\begin{align}
    Q(u_t;\phi_u) = (1-\phi_z)^{1 - \frac{u_t + 1}{2}} \phi_z^{ \frac{u_t + 1}{2}} 
\end{align}

From Equation \eqref{eq:appB_vfe_problem_qu} it follows that the optimal $\phi_u$ is that which minimizes the Kullback-Leibler divergence between the approximate posterior $Q(u_t; \phi_u)$ and the prior $P(u_t)$, which is a softmax function of the expected free energy of actions $\mathbf{G}(u_t)$. In this particular generative model, the expected free energy can be written as a single term that scores the `expected utility' of each action \cite{friston2015active, friston2017active}:

\begin{align}
    \mathbf{G}(u_t) &= -\mathbb{E}_{Q(\mathcal{O}_{t+1}|u_t)}[\ln \tilde{P}(\mathcal{O}_{t+1})]
    \label{eq:appB_efe_utility}
\end{align}

To compute this, we need to compute the `variational marginal' over $\mathcal{O}_{t+1}$, denoted $Q(\mathcal{O}_{t+1}|u_t)$:

\begin{align}
    Q(\mathcal{O}_{t+1}|u_t) &= \mathbb{E}_{Q(z_t;\phi^{*}_z)}[P(\mathcal{O}_{t+1}|z_t, u_t)]
\end{align}

We can simplify the expression for $Q(\mathcal{O}_{t+1}|u_t)$ when we take advantage of the Bernoulli-parameterization of the posterior over hidden states $Q(z;\phi^{*}_z)$. This allows us to then write the variational marginals, conditioned on different actions as a matrix, with one column for each setting of $u_t$:

\begin{align}
    Q(\mathcal{O}_{t+1}|u_t) &= \begin{bmatrix}\phi^{*}_z  & 1 - \phi^{*}_z \\ 1 - \phi^{*}_z & \phi^{*}_z  \end{bmatrix} \label{eq:appB_qo_bernoulli_matrix}
\end{align}

The expected utility (and thus the negative expected free energy) is then computed as the dot-product of each column of the matrix expressed in Equation \eqref{eq:appB_qo_bernoulli_matrix} with the log of the prior preferences $\tilde{P}(\mathcal{O}_{t+1})$:

\begin{align}
     \mathbb{E}_{Q(\mathcal{O}_{t+1}|u_t)}[\ln \tilde{P}(\mathcal{O}_{t+1})] &= \begin{bmatrix} -\infty \phi^{*}_z \\ -\infty (1 - \phi^{*}_z)\end{bmatrix} \notag \\
     \implies \mathbf{G}(u_t) &= \begin{bmatrix} \infty \phi^{*}_z \\ \infty (1 - \phi^{*}_z) \end{bmatrix}
\end{align}

Because the probability of an action is proportional to its negative expected free energy, this allows us to write the Bernoulli parameter $\phi^{*}_u$ of the posterior over actions directly in terms of the parameter of the state posterior:

\begin{align}
    \phi^{*}_u &= \frac{1}{1 + \exp(\beta (\infty ( 1- \phi^{*}_z)))} \notag \\
    &= \frac{1}{1 + C\exp(-\phi^{*}_z)))}
    \label{eq:appB_phi_z_to_phi_u}
\end{align}

The inverse temperature parameter $\beta$ is an arbitrary re-scaling factor that can be used to linearize the sigmoid function in \eqref{eq:appB_phi_z_to_phi_u} over the range $[ 0, 1 ]$ such that

\begin{align}
    \phi^{*}_u &\approx \phi^{*}_z \label{eq:appB_probability_matching}
\end{align}

Note that the equivalence relation in Equation \eqref{eq:appB_probability_matching} is only possible due to the infinite precisions $\omega$ and $\xi$ of the likelihood and prior distributions over the `optimality' variable $P(\mathcal{O}_{t+1}|u_t, z_t)$ and $\tilde{P}(\mathcal{O}_{t+1})$, and from an appropriately re-scaled $\beta$ parameter that linearizes the sigmoid relationship in Equation \eqref{eq:appB_phi_z_to_phi_u}.

\subsection{Action sampling as probability matching}

Now that we have an expression for the parameter $\phi^{*}_u$ of the posterior over control states $Q(u_t; \phi^{*}_u)$, an agent can generate a spin state by simply sampling from this posterior over actions:

\begin{align}
    \sigma &\sim Q(u_t;\phi^{*}_u) \notag \\
    &\sim Q(z_t;\phi^{*}_z) \triangleq P(z_t|\tilde{\sigma};\gamma, \zeta)
\end{align}

In short, each agent samples its spin state from a posterior belief over the state of the latent variable $z_t$, rendering their action-selection a type of probability matching \cite{vulkan2000economist, shanks2002re, gaissmaier2008smart}, whereby actions (whether to spin `UP' or `DOWN') are proportional to the probability they are assigned in the agent's posterior belief. Each agent's sampled spin state also serves as an observation ($\sigma_j$ for some $j$) for the other agents that the focal agent is a neighbour of. This collective active inference scheme corresponds to a particular form of sampling from the stationary distribution of a spin glass model known as Glauber dynamics \cite{glauber1963time}. Crucially, however, the temporal scheduling of the action-updating across the group determines which stationary distribution the system samples from. We explore this distinction in the next section.

\section{Temporal scheduling of action sampling}\label{sec:appendix_C}

In this appendix we examine how the stationary distribution from which the collective active inference system samples depends on the order in which actions are updated across all agents in the network. First, we consider the case of synchronous action updates (all agents update their actions in parallel and only observe the- spin states of their neighbours from the last timestep), and show how this system samples from a different stationary distribution than the one defined by the standard Ising energy provided in Equation \eqref{eq:ising_energy_func}. We then derive the more `classical' case of asynchronous updates, where agents update their spins one at a time, and show how in this case the system converges to the standard statioanry distribution of the Ising model. This Appendix thus explains one of the `fragilities' mentioned in the main text, that threaten the unique equivalence between local active inference dynamics and a unique interpretation at the global level in terms of inference.

We denote some agent's spin using $\sigma_i$ and its set of neighbours as $M_i$. The local sum of spins or spin difference $\sum_{j \in M} \sigma_j$ for agent $i$ we denote $\Delta_i \sigma = \sum_{j \in M_i} \sigma_j$.

\subsection{Synchronous updates}

To derive the stationary distribution in case of synchronous updates, we can take advantage of the following detailed balance relation, which obtains in the case of systems at thermodynamic equilibrium:

\begin{align}
    \frac{P(\tilde{\sigma})}{P(\tilde{\sigma}')} &= \frac{P(\tilde{\sigma}|\tilde{\sigma}')}{P(\tilde{\sigma}'|\tilde{\sigma})} \notag \\
    \implies P(\tilde{\sigma}) &= \frac{P(\tilde{\sigma}|\tilde{\sigma}')P(\tilde{\sigma}')}{P(\tilde{\sigma}'|\tilde{\sigma})} \label{eq:appC_detailed_balance_cond}
\end{align}
where $\tilde{\sigma}$ and $\tilde{\sigma}'$ are spin configurations at two adjacent times $\tau$ and $\tau+1$. In the case of synchronous updates (all spins are sampled simultaneously, given the spins at the last timestep), then the spin action of each agent $\sigma_i'$ at time $\tau+1$ is conditionally independent of all other spins, given the vector of spins $\tilde{\sigma}$ at the previous timestep $\tau$. We can therefore expand the `forward' transition distribution $P(\tilde{\sigma}'|\tilde{\sigma})$ as a product over the action posteriors of each agent:

\begin{align}
    P(\tilde{\sigma}'|\tilde{\sigma}) &= P(\sigma_1|\tilde{\sigma})P(\sigma_1|\tilde{\sigma})...P(\sigma_N|\tilde{\sigma}) \notag \\
    &= \prod_i Q(u_t; \phi^{*}_{u, i}) \notag \\
    &= \prod_{i} \frac{\exp\left( \sigma'_i \left(\zeta + \gamma \sum_{j \in M_i} \sigma_j \right)\right)}{2\cosh\left(\zeta + \gamma \sum_{j \in M_i} \sigma_j \right)} \notag \\
    &= \exp\left(\sum_i \sigma'_i (\zeta + \gamma \sum_{j \in M_i} \sigma_j ) - \sum_i\log\left(2 \cosh (\zeta + \gamma \sum_{j \in M_i} \sigma_j )\right) \right)
\end{align}

Note we have replaced each latent variable in the posterior $z$ with the agent's own spin state $\sigma_i$, because there is a one-to-one mapping between the posterior over $z_t$ and the posterior over actions $\sigma_i$.

The reverse transition distribution, yielding the probability of transitioning from configuration $\tilde{\sigma}' \to \tilde{\sigma}$ is the same expression as for the forward transition, except that $\sigma_i'$ and $\sigma_i$ are swapped:

\begin{align}
    P(\tilde{\sigma}|\tilde{\sigma}')&= \exp\left(\sum_i \sigma_i (\zeta + \gamma \sum_{j \in M_i} \sigma'_j ) - \sum_i\log\left(2 \cosh (\zeta + \gamma \sum_{j \in M_i} \sigma'_j)\right) \right)
\end{align}

The detailed balance equation in \eqref{eq:appC_detailed_balance_cond} then tells us that the stationary probability distribution over $\tilde{\sigma}$ is proportional to the ratio of the backwards transition to the forwards transition:

\begin{align}
    \frac{P(\tilde{\sigma})}{P(\tilde{\sigma'})} &= \frac{\exp\left(\sum_i \sigma_i (\zeta + \gamma \sum_{j \in M_i} \sigma'_j ) - \sum_i\log\left(2 \cosh (\zeta + \gamma \sum_{j \in M_i} \sigma'_j)\right) \right)}{\exp\left(\sum_i \sigma'_i (\zeta + \gamma \sum_{j \in M_i} \sigma_j ) - \sum_i\log\left(2 \cosh (\zeta + \gamma \sum_{j \in M_i} \sigma_j )\right) \right)} \notag \\
    &= \frac{\exp\left(\zeta \sum_i \sigma_i + \gamma \sum_{\langle i, j\rangle} \sigma_i \sigma_j'\right) \exp\left(-\sum_i \log\left(2 \cosh(\zeta + \gamma \sum_{j \in M_i} \sigma_j'\right)\right)}{\exp\left(\zeta \sum_i \sigma_i' + \gamma \sum_{\langle i, j\rangle} \sigma_i' \sigma_j\right) \exp\left(-\sum_i \log\left(2 \cosh(\zeta + \gamma \sum_{j \in M_i} \sigma_j\right)\right)} \notag \\
    &= \frac{\exp\left(\zeta \sum_i \sigma_i  +\sum_i \log \left(2 \cosh (\zeta + \gamma \sum_{j \in M_i} \sigma_j\right)\right)}{\exp\left(\zeta \sum_i  \sigma_i' + \sum_i \log \left(2 \cosh (\zeta + \gamma \sum_{j \in M_i} \sigma_j'\right)\right)}
\end{align}

Therefore, we can write down the stationary distribution in the case of synchronous updates as an exponential term normalized by a partition function:

\begin{align}
\label{eq:stationary_dist_synchronous}
    P(\tilde{\sigma}) &= Z^{-1}\exp\left(\zeta \sum_i \sigma_i  + \sum_i \log \left(2 \cosh (\zeta + \gamma \sum_{j \in M_i} \sigma_j)\right)\right) \notag\\
    Z &= \sum_{\tilde{\sigma}} \exp\left(\zeta \sum_i \sigma_i  + \sum_i \log \left(2 \cosh (\zeta + \gamma \sum_{j \in M_i} \sigma_j)\right)\right)
\end{align}

Note that the action update for an individual agent can still be written in terms of the local energy difference $\Delta_i E$, where the energy is defined using the standard Hamiltonian function given by Equation \eqref{eq:ising_energy_func} in the main text. However, due to the temporal sampling of each agent's action with respect to the others, the system collectively sample from a system with a different energy function and Gibbs measure, given by Equation \eqref{eq:stationary_dist_synchronous}. This energy function is therefore nonlinear and can be written:

\begin{align}
    E_{sync}(\tilde{\sigma}) &= -\zeta \sum_i \sigma - \sum_i \log (2\cosh(\zeta + \gamma \sum_{j \in M_i} \sigma_j))
\end{align}

\subsection{Asynchronous updates}

Now we treat the case where agents update their agents one-by-one or asynchronously. This means that at each timestep only one agent is updated, and that particular agent uses the spin states of all the other agents at the last timestep as inputs for its posterior inference.

We can write down the forward transition as follows, using the notation $\sigma_{\setminus i}$ to denote all the spins except for $\sigma_i$:

\begin{align}
    p(\sigma_i',\tilde{\sigma}_{\setminus i}| \tilde{\sigma}) &= \frac{\exp(\sigma_i'(\zeta + \gamma \sum_{j \in M_i} \sigma_j))}{2 \cosh (\zeta + \gamma \sum_{j \in M_i} \sigma_j)}
\end{align}
which indicates that only agent $i$ is updated at the current timestep. The detailed balance condition implies that 

\begin{align}
    p(\sigma_i',\tilde{\sigma}_{\setminus i}| \tilde{\sigma}) p(\tilde{\sigma}) = p(\tilde{\sigma}|\sigma_i',\tilde{\sigma}_{\setminus i}) p(\sigma_i',\tilde{\sigma}_{\setminus i})
\end{align}

Then

\begin{align}
    \frac{p(\tilde{\sigma})}{p(\sigma_i',\tilde{\sigma}_{\setminus i})} =& \frac{p(\tilde{\sigma}|\sigma_i',\tilde{\sigma}_{\setminus i})}{p(\sigma_i',\tilde{\sigma}_{\setminus i}| \tilde{\sigma})} = \frac{\exp(\sigma_i(\zeta + \gamma \sum_{j \in M_i} \sigma_j) - \log (2\cosh(\zeta + \gamma \sum_{j \in M_i} \sigma_j)))}{\exp(\sigma_i'(\zeta + \gamma \sum_{j\in M_i} \sigma_j) - \log (2\cosh(\zeta + \gamma\sum_{j\in M_i}\sigma_j)))}
    \\ =& \frac{\exp(\sigma_i(\zeta +  \gamma\sum_{j\in M_i} \sigma_j))}{\exp(\sigma_i'(\zeta + \gamma \sum_{j\in M_ii} \sigma_j ))}
\end{align}

By repeating this operation for every agent (i.e. $N-1$ more times), then we arrive at:

\begin{align}
   \frac{ p(\tilde{\sigma})}{p(\tilde{\sigma}')} = \frac{ p(\tilde{\sigma})}{p(\sigma_i',\tilde{\sigma}_{\setminus i})} \frac{p(\sigma_i',\tilde{\sigma}_{\setminus i})}{p(\sigma_i',\sigma_j',\tilde{\sigma}_{\setminus i,j})} \ldots \frac{p(\tilde{\sigma}'_{\setminus i},\sigma_i)}{p(\tilde{\sigma}')}  =& \frac{\exp(\zeta \sum_i \sigma_i +  \gamma\sum_{i<j} \sigma_i  \sigma_j )}{\exp (\zeta \sum_i \sigma'_i + \gamma \sum_{i<j}\sigma'_i \sigma'_j)} \label{eq:appC_async_update_derivation}
\end{align}

We can therefore write the marginal distributions  $p(\tilde{\sigma})$ as proportional to the numerator  of the last term in Equation \eqref{eq:appC_async_update_derivation}\footnote{Note that because of assumption that the system is at thermal equilibrium, the same reasoning could be applied to write the distribution over $p(\tilde{\sigma}')$ in terms of the denominator of Equation \eqref{eq:appC_async_update_derivation}}:

\begin{align}
    p(\tilde{\sigma}) &\propto \exp(\zeta \sum_i \sigma_i + \gamma \sum_{\langle i, j\rangle} \sigma_i \sigma_j) \notag \\
    \implies p(x) =& Z^{-1} \exp(\zeta \sum_i \sigma_i + \gamma \sum_{\langle i, j\rangle}  \sigma_i \sigma_j)
\end{align}

We thus recover the original stationary distribution with the standard, linear energy function as given by Equation \eqref{eq:ising_energy_func} in the main text, written now in terms of generative model parameters $\gamma, \zeta$ instead of the standard `couplings' and `biases' $J, h$:

\begin{align}
    \label{eq:energy_function_async}
    E_{async}(\tilde{\sigma}) &= -\gamma \sum_{\langle i, j \rangle}\sigma_i \sigma_j - \zeta \sum_i \sigma_i
\end{align}

\printbibliography[title={Supplemental References}]
\end{refsection}

\end{document}